\documentclass[conference]{IEEEtran}
\usepackage{cite}
\ifCLASSINFOpdf
\else
\fi
\usepackage[linesnumbered,vlined]{algorithm2e}
\usepackage{epsfig}
\usepackage{amsfonts}
\usepackage{amssymb}
\usepackage{subfigure}
\usepackage[usenames]{color}
\usepackage{soul}
\usepackage{comment}
\usepackage{multirow}
\usepackage[cmex10]{amsmath}
\usepackage{bm}
\usepackage{setspace}

\newcommand{\Az}[1]{{\color{black}{#1}}}
\newcommand{\AzCom}[1]{}
\newcommand{\AzDel}[1]{}

\begin{document}
\title{Future Evolution of CSMA Protocols for the IEEE 802.11 Standard}

% author names and affiliations
% use a multiple column layout for up to three different
% affiliations
\author{\IEEEauthorblockN{Luis Sanabria-Russo, Azadeh Faridi, Boris Bellalta, Jaume Barcelo, Miquel Oliver}
\IEEEauthorblockA{
Universitat Pompeu Fabra\\
Roc Boronat 183, 08018 Barcelona\\
Email: \{name.surname@upf.edu\}}
}

% use for special paper notices
%\IEEEspecialpapernotice{(Invited Paper)}

% make the title area
\maketitle

\begin{abstract}
%\boldmath
In this paper a candidate protocol to replace the prevalent CSMA/CA medium access control in Wireless Local Area Networks is presented. The proposed protocol can achieve higher throughput than CSMA/CA, while maintaining fairness, and without additional implementation complexity. Under certain circumstances, it is able to reach and maintain collision-free operation, even when the number of contenders is variable and potentially large. It is backward compatible, allowing for new and legacy stations to coexist without degrading one another's performance, a property that can make the adoption process by future versions of the standard smooth and inexpensive\footnotemark[1]\footnotetext[1]{This paper has been accepted in the Second IEEE ICC Workshop 2013 on Telecommunication Standards: From Research to Standards.}.

% In this paper we have described a simply implementable protocol that can be used as a replacement of the prevalent CSMA/CA protocol used in the IEEE 802.11 standard. Compared to the CSMA/CA protocol, the proposed protocol, called CSMA/ECA with hysteresis and fair-share, offers performance boost in terms of throughput while preserving long-term fairness. In fact, under certain conditions, collision-free operation can be reached in a distributed and adaptive manner, even as the number of nodes in the network increase. Furthermore, this protocol is backward compatible, and can operate with nodes using the legacy protocol without degrading the performance of the legacy nodes. All these properties make the proposed protocol a very suitable candidate to replace CSMA/CA in the upcoming revisions of the standard.

% In this paper we present the requirements of candidate protocols to replace the prevalent CSMA/CA medium access control.
% We discuss the possibility of further preventing collisions and provide an overview of the related work.
% We specify a protocol that is a candidate for replacing CSMA/CA in pseudo-code and use simulation to assess performance metrics such as throughput, fairness and collision probability.
\end{abstract}

% For peer review papers, you can put extra information on the cover
% page as needed:
% \ifCLASSOPTIONpeerreview
% \begin{center} \bfseries EDICS Category: 3-BBND \end{center}
% \fi
%
% For peerreview papers, this IEEEtran command inserts a page break and
% creates the second title. It will be ignored for other modes.
\IEEEpeerreviewmaketitle

\section{Introduction}
% no \IEEEPARstart
% You must have at least 2 lines in the paragraph with the drop letter
% (should never be an issue)

Since the popularization of the IEEE 802.11 standard, several works (e.g.,  \cite{bharghavan1994map,wang2004ncr,cali2000dti,lopez-toledo2006aoi,
barcelo2008lba,bellalta2009vtc,he2009srb,barcelo2010fcc,fang2011dlm,hui2011epp,barcelo2011tcf}) have proposed modifications to CSMA/CA (Carrier Sense Multiple Access with Collision Avoidance), the contention protocol used in the standard for controling the access to the shared medium. {Despite the throughput improvement offered by all these works, none of the proposed modifications have yet been adopted by the standard.}

A suitable candidate to replace CSMA/CA, should
\begin{itemize}
  \item provide performance advantages in terms of throughput and/or short-term fairness;
  \item be backward compatible with the current implementation;
  \item support a large number of simultaneous contenders; and
  \item be a simple evolution, in terms of implementation, to ease the transition and reduce the time to market (optional but desirable).
\end{itemize}

The aforementioned works can be categorized in three groups regarding the approach they use. {In this paper, we will focus on the last of these three groups, proposing an additional enhancement to make it more adaptable to network size variations, and make the case for its adoption by the future versions of the standard.}

In the first group, represented here by \cite{bharghavan1994map,wang2004ncr}, a throughput increase is obtained in saturation conditions by preventing the contention window from resetting to its minimum value after each successful transmission. However this performance gain comes at the expense of reduced short-term fairness, since nodes that have recently gone through a sequence of consecutive collisions are forced to stay at a higher backoff stage and thus are further penalized by less frequent transmissions.

In the second group of works, exemplified here by \cite{cali2000dti,lopez-toledo2006aoi}, an accurate estimate of the number of contenders is used to adjust the contention parameters. This approach offers some throughput and fairness gains, but at the expense of increased implementation complexity. In addition, as the number of contenders is estimated relying on the number of collisions, the presence of channel errors renders the estimation inaccurate.
Furthermore, there is a fundamental trade-off between the stability and the speed of reaction of the estimation.

When it comes to backward compatibility, neither of the aforementioned solutions are able to fairly share the medium with legacy devices.
In fact, these proposals are, generally speaking, less aggressive than the currently implemented CSMA/CA protocol. Consequently, in a hypothetical mixed network in which the new and old protocols coexist, the new stations receive a smaller share of the available bandwidth.

A more important limitation of the two approaches exposed so far is that the throughput is bounded by that of CSMA/CA with optimal configuration \cite{cali2000dti,bianchi2000pai}.

The third group of solutions, which is the focus of the current paper, can deliver a throughput above the maximum attainable by CSMA/CA. This performance boost is achieved mainly by the use of deterministic backoff after successful transmissions, which reduces the chances of collisions for nodes that were successful in their previous transmission. Furthermore, under certain conditions, a collision-free operation can be reached. {We will refer hereafter to the class of algorithms that use deterministic backoffs after successful transmissions as CSMA with Enhanced Collision Avoidance (CSMA/ECA).} This approach was first introduced in \cite{barcelo2008lba}, and later a more detailed analysis for both saturated and unsaturated conditions was presented in \cite{bellalta2009vtc}.
A more in-depth study is carried out in \cite{he2009srb}, including realistic elements such as the possibility of Clear Channel Assessment errors.
Different aspects of fairness are addressed in \cite{he2009srb,barcelo2010fcc,fang2011dlm}.
The performance in realistic channels, taking into account the Auto Rate Fallback mechanism, is evaluated in \cite{martorell2012pec,martorell2012tfl}.

Quality-of-service and traffic differentiation in CSMA/ECA is discussed in \cite{barcelo2009tpc, he2009srb}.
The performance of CSMA/ECA in the presence of channel errors is considered in \cite{he2009srb, barcelo2010fcc, fang2011dlm, barcelo2012mdc}.
Other aspects such as coexistence with legacy stations and the performance when nodes can arbitrarily enter and leave the network have been covered in \cite{he2009srb, barcelo2010fcc, fang2011dlm}.
Results for the impact of slot drift (non-ideal clocks) in protocols similar to CSMA/ECA can be found in \cite{gong2012asd}.

Even though initial research efforts where focused on the WLAN collision problem, more recent works try to extend the idea to multi-hop networks.
In \cite{hui2011epp}, the multi-hop slotted case is explored.
A solution that does not require network-wide slot synchronization is studied in  \cite{barcelo2013dcc}.

\AzCom{Greenfield paragraph moved to the Results section} The same principles that we exploit to prevent collisions in WLANs can be used in other areas of radio resource management in wireless area networks \cite{duffy2011dcs,checco2012scs,checco2012lbc,khan2013aso}.

In all the previous work on collision-free operation in WLANs mentioned so far, there is the limitation that the number of contenders should not exceed the value of the deterministic backoff used after successful transmissions.
If this value is exceeded, it is no longer possible to achieve collision-free operation.
A first solution to solve this problem is presented in \cite{barcelo2011tcf}, but it requires the presence of a central entity (typically the access point) that instructs the other nodes to adjust the value of their deterministic backoff.

In the current paper, we present a completely distributed solution, called {\it backoff stage hysteresis} or simply hysteresis, which is a simple modification to CSMA/ECA to accommodate a large number of contenders in an adaptive and distributed fashion. Hysteresis by itself, comes at the expense of reduction in long-term fairness. Using hysteresis together with what we will refer to as {\it fair-share} allows for reaching collision-free operation with self-adaptation to a variable, and potentially large, number of contenders while preserving fairness. The concept of fair-share was first introduced in \cite{fang2011dlm}. We will furthermore detail pseudo-code algorithms for three different types of CSMA/ECA (basic, with hysteresis, and with both hysteresis and fair-share) and highlight how they can be easily implemented using simple modifications to the currently-used CSMA/CA algorithm. We will then present some performance evaluation, quantifying the performance gain that these three types of CSMA/ECA can achieve, both in terms of throughput and fairness.

Note that the coexistence and fairness of basic CSMA/ECA and CSMA/CA has been studied in \cite{barcelo2010fcc}.
The addition of hysteresis and fair-share to the basic CSMA/ECA may represent less frequent transmissions and more packets per transmission
These two modifications do not interfere in any way with the execution of the legacy protocol; therefore, CSMA/CA with hysteresis and fair-share can also coexist with the currently deployed protocol.

\section{Enhanced CSMA/CA}

In this section and the next, we make the case for the suitability of CSMA/ECA with hysteresis and fair-share as a future replacement to the CSMA/CA protocol. To do this, we begin by presenting a simple pseudo-code which captures the essence of the CSMA/CA protocol as it is currently implemented in the IEEE 802.11 standard. We will then show how CSMA/ECA can be implemented by a minor modification to the CSMA/CA algorithm. We then describe hysteresis and fair-share, by presenting minor modifications, one by one, to the CSMA/ECA algorithm. The final product---CSMA/ECA with hysteresis and fair-share---satisfies the four requirements specified in the introduction, which makes it a suitable candidate to replace CSMA/CA in the upcoming revisions of the standard. We would like to emphasize here that the basic CSMA/ECA and the concept of fair-share are not presented here for the first time, however, we include the algorithms corresponding to these protocols for both completeness and ease of comparison between protocols.

Algorithm \ref{alg:CSMA_CA} describes the CSMA/CA protocol that is used in current networks.
When a station joins the contention, it initializes the retry attempt counter $r$ and the backoff stage $s$ to zero. The backoff counter $b$ is initialized using a uniform random distribution and the minimum contention window $\text{CW}_{\min}$. After each collision, the retry attempt counter and the backoff stage counter are incremented. As a consequence of the incremented backoff stage, a larger contention window is used. Note that there is a maximum backoff stage $S$ and a maximum retry limit $R$ specified by the protocol. When the number of transmission attempts on a packet reaches the maximum retry limit, the packet is discarded. Furthermore, $r$ and $s$ are reset to zero, and a new value for $b$ is computed both when the retry limit is reached and when the transmission is successful. Note that the pseudo-code performs the same action in lines $18$ and $21$. Therefore, the algorithm could be further simplified if instead this action were moved to immediately after line $16$. However, the current presentation of the algorithm eases the comparison between the CSMA/CA and the enhancements that follow.

% In this section, we describe the original CSMA/CA protocol and then we describe each of the modifications that we propose, one by one.
%
%
% In this section we describe the CSMA/CA protocol as it is currently implemented in the IEEE 802.11 and an evolution of the standard that satisfies the four requirements specified in the introduction.
% Therefore, the presented protocol is a valid candidate to replace CSMA/CA in the upcoming revisions of the standard.

There are three changes in the CSMA/ECA with hysteresis and fair-share compared to the legacy CSMA/CA protocol. Firstly, a deterministic backoff is used after successful transmissions. This simple modification converts CSMA/CA into basic CSMA/ECA.
Secondly, the backoff stage is not reset after a packet is serviced.
The backoff stage is reset only when the station leaves the contention because it has no packet to be transmitted. This modification is what we have been referring to as hysteresis.
And thirdly, the number of packets transmitted in every transmission attempt is chosen as a function of the backoff stage, which is what we call fair-share.
Note that current standards already support the transmission of multiple packets in a single slot.

Algorithm \ref{alg:CSMA_ECA} describes the basic CSMA/ECA in which a deterministic backoff is used after successful transmissions. The only change with respect to CSMA/CA is in fact in line $18$, where the random assignment of $b$ in CSMA/CA is replaced by a deterministic assignment in CSMA/ECA. Note that at this point the value of $s$ is zero, and therefore the assigned deterministic value is in fact $b = \text{CW}_{\min}/2-1$. This value is roughly equal to the expectation of the backoff chosen in line $18$ in Algorithm \ref{alg:CSMA_CA}. This particular choice improves fairness between new stations and legacy stations. \AzCom{Does it? Even if they are not using aggregation?}

\begin{algorithm}[ht!!!]
\While{the device is on}
{
  %\tcc{Initialize retransmission attempts.}
  $r \leftarrow 0$; $s \leftarrow 0$\;
  %\tcc{Initialize backoff counter.}
  $b \leftarrow \mathcal{U}[0,2^s{\rm{CW}_{min}}-1]$\;
  \While{there is a packet to transmit}{
    %\tcc{Initialize $a$.}
    \Repeat{($r = R$) or (success)}{
      %\tcc{First, backoff.}
      \While{$b>0$}{
        wait 1 slot\;
        $b \leftarrow b-1$\;
      }
      \colorbox{yellow}{Attempt transmission of 1 packet;}\\
      \If{collision}{
        %\tcc{Random backoff.}
        $r \leftarrow r+1$\;
        $s \leftarrow \min (s+1,S)$\;
        $b \leftarrow \mathcal{U}[0, 2^s {\rm{CW}_{min}} -1]$\;
      }
    }
    $r \leftarrow 0$\;
    \colorbox{yellow}{$s \leftarrow 0$;}\\
	 \eIf{success}{
      %\tcc{Random backoff.}
      \colorbox{yellow}{$b \leftarrow \mathcal{U}[0,2^{s}{\rm{CW}_{min}}-1]$;}\\
    }
    {
      Discard packet\;
      $b \leftarrow \mathcal{U}[0,2^s {\rm{CW}_{min}}-1]$\;
    }
  }
  Wait until there is a packet to transmit\;
}
\caption{CSMA/CA}
\label{alg:CSMA_CA}
\end{algorithm}

\begin{algorithm}[ht!!!]
\While{the device is on}
{
  $r \leftarrow 0$ ; $s \leftarrow 0$\;
  $b \leftarrow \mathcal{U}[0,2^s{\rm{CW}_{min}}-1]$\;
  \While{there is a packet to transmit}{
    %\tcc{Initialize $a$.}
    \Repeat{($r = R$) or (success)}{
      %\tcc{First, backoff.}
      \While{$b>0$}{
        wait 1 slot\;
        $b \leftarrow b-1$\;
      }
      \colorbox{yellow}{Attempt transmission of 1 packet;}\\
      \If{collision}
      {
        %\tcc{Random backoff.}
        $r \leftarrow r+1$\;
        $s \leftarrow \min (s+1,S)$\;
        $b \leftarrow \mathcal{U}[0, 2^s {\rm{CW}_{min}} -1]$\;
      }
    }
    $r \leftarrow 0$\;
    \colorbox{yellow}{$s \leftarrow 0$;}\\
    \eIf{success}{
      %\tcc{Random backoff.}
      \colorbox{yellow}{$b \leftarrow (2^{s}{\rm{CW}_{min}})/2-1$;}\\
    }
    {
      Discard packet\;
      $b \leftarrow \mathcal{U}[0, 2^s {\rm{CW}_{min}}-1]$\;
    }
  }
  Wait until there is a packet to transmit\;
}
\caption{Basic CSMA/ECA}
\label{alg:CSMA_ECA}
\end{algorithm}

\begin{algorithm}[ht!!!]
\While{the device is on}
{
  $r \leftarrow 0$ ; $s \leftarrow 0$\;
  $b \leftarrow \mathcal{U}[0,2^s{\rm{CW}_{min}}-1]$\;
  \While{there is a packet to transmit}{
    %\tcc{Initialize $a$.}
    \Repeat{($r = R$) or (success)}{
      %\tcc{First, backoff.}
      \While{$b>0$}{
        wait 1 slot\;
        $b \leftarrow b-1$\;
      }
      \colorbox{yellow}{Attempt transmission of 1 packet;}\\
      \If{collision}{
        %\tcc{Random backoff.}
        $r \leftarrow r+1$\;
        $s \leftarrow \min (s+1,S)$\;
        $b \leftarrow \mathcal{U}[0, 2^s {\rm{CW}_{min}} -1]$\;
      }
    }
    $r \leftarrow 0$\;
    %$s \leftarrow 0$\;
    \eIf{success}{
      %\tcc{Random backoff.}
      \colorbox{yellow}{$b \leftarrow (2^{s}{\rm{CW}_{min}})/2-1$;}\\
    }
    {
      Discard packet\;
      $b \leftarrow \mathcal{U}[0, 2^s {\rm{CW}_{min}}-1]$\;
    }
  }
  Wait until there is a packet to transmit\;
}
\vspace{0.2cm}
\caption{CSMA/ECA with hysteresis}
\label{alg:CSMA_ECA_hist}
\end{algorithm}

\begin{algorithm}[ht!!!]
\While{the device is on}
{
  $r \leftarrow 0$ ; $s \leftarrow 0$\;
  $b \leftarrow \mathcal{U}[0,2^s\rm{CW}_{min}-1]$\;
  \While{there is a packet to transmit}{
    %\tcc{Initialize $a$.}
    \Repeat{($r = R$) or (success)}{
      %\tcc{First, backoff.}
      \While{$b>0$}{
        wait 1 slot\;
        $b \leftarrow b-1$\;
      }
      \colorbox{yellow}{Attempt transmission of $2^s$ packets;}\\
      \If{collision}{
        %\tcc{Random backoff.}
        $r \leftarrow r+1$\;
        $s \leftarrow \min (s+1,S)$\;
        $b \leftarrow \mathcal{U}[0, 2^s  \rm{CW}_{min} -1]$\;
      }
    }
    $r \leftarrow 0$\;
    %$s \leftarrow 0$\;
    \eIf{success}{
      %\tcc{Random backoff.}
      \colorbox{yellow}{$b \leftarrow (2^{s}\rm{CW}_{min})/2-1$;}\\
    }
    {
      Discard packet\;
      $b \leftarrow \mathcal{U}[0, 2^s \rm{CW}_{min}-1]$\;
    }
  }
  Wait until there is a packet to transmit\;
}
\vspace{0.2cm}
\caption{CSMA/ECA with hysteresis and fair-share}
\label{alg:CSMA_ECA_hist_fair}
\end{algorithm}

Hysteresis is included in Algorithm \ref{alg:CSMA_ECA_hist}. Adding hysteresis is as simple as removing line $16$ from Algorithm \ref{alg:CSMA_ECA}. This means that the deterministic backoff value chosen after a successful transmission in this case is half the length of the contention window of the backoff stage in which the successful transmission has occurred. In fact, the backoff stage is reset only when the node has no packet to serve. This allows for the use of larger deterministic backoff values, which means that the network can reach a collision-free operation for a higher number of nodes. Since in the collision-free operation, some nodes may have higher backoff values than others, this protocol does not share the medium in a fair manner among nodes.

Finally, fair-share is implemented in Algorithm \ref{alg:CSMA_ECA_hist_fair} by simply modifying
line $9$ of Algorithm \ref{alg:CSMA_ECA_hist} to increase the number of transmitted packets from one to $2^s$. This way, at each transmission opportunity in collision-free operation, each node transmits a number of packets proportional to the length of its deterministic backoff. Therefore, nodes that have to wait longer between transmission opportunities get to transmit more packets, hence fairness is preserved.

\begin{figure*}[!!!htb]
\centering
\includegraphics[width=7.0in]{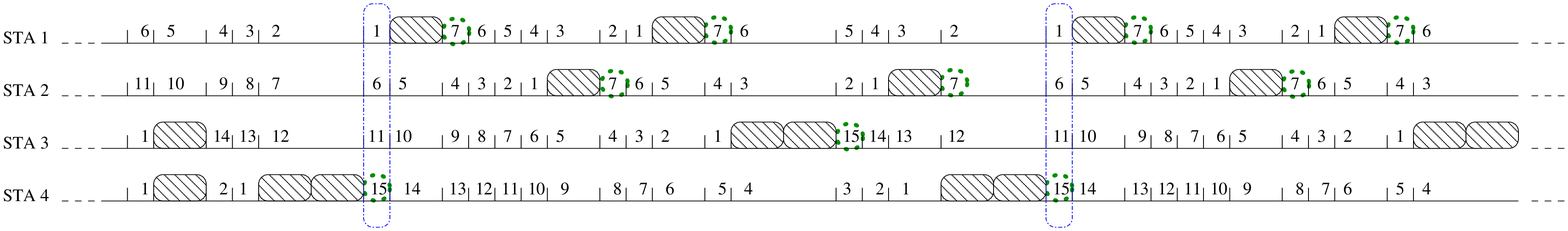}
\caption{An example of contention among CSMA/ECA stations with hysteresis and fair-share.}
\label{fig:csma_eca_different_backoff}
\end{figure*}

An example of the operation of CSMA/ECA with hysteresis is presented in Fig. \ref{fig:csma_eca_different_backoff} for a network of $4$ contending stations.
For each station the backoff counter value at each station is indicated in every slot and the transmissions are represented by rounded rectancles. \Az{In the first slot shown, all stations are in the first backoff stage ($s = 0$, $\text{CW}_{\min} = 16$) and have chosen a random backoff value $b \sim \mathcal{U}[0,15]$. Stations $3$ and $4$ happen to have the same backoff counter value, and when their counter reaches zero, their transmissions collide. Therefore, they increase their backoff stage ($s = 1$) and select another random backoff value accordingly ($b \sim \mathcal{U}[0,31]$). On the other hand, stations $1$ and $2$ manage to successfully transmit their packets and move to the deterministic behavior for their next transmission, both selecting a deterministic value equal to $7 = 2^s \rm{CW}_{min}/2 - 1$. The first slot in which all stations have different backoff values, marked in the figure with a blue dashed line, can be viewed as the beginning of the collision-free operation. Since in this case no two stations will transmit simultaneously, there will be no collision and, by switching to the deterministic behavior, they will all periodically transmit in the same relative position. Note that stations $3$ and $4$ move to deterministic behavior when they are in backoff stage $s=1$. Therefore, they both use $15 = 2^s \rm{CW}_{min}/2 - 1$ as their deterministic backoff value and send two packets at each transmission opportunity (fair-share). The length of the deterministic cycle for the network in this example is $32$ slots, the least common multiple of all stations' cycles.}
% The stations initially pick a random backoff.}
% Those stations with the same backoff value collide and choose a new backoff value.
% When all the stations have a different backoff value, no more collisions will occur.
% At this point the behaviour of the system is cyclic and deterministic.
% Note that the state of the system at the beginning of the cycle is the backoff counter of the four stations: 4, 9, 14 and 2.
% It is exactly the same state that the first slot of the next cycle.

% The length of the cycle (32 slots) is indicated by the arrows at the bottom of the figure.
% The stations that collide double their contention window, the deterministic backoff after successes and the number of transmitted packet in each transmission attempt.

\section{Performance Evaluation} \label{sec:perf_eval}

%Ok, I'd use CWmin/2 instead of Nmax, and "s" instead of "l" for the backoff stage.

%\AzDel{In the present workshop paper, we focus on a simple greenfield scenario with
%a single traffic class, ideal channel, ideal clocks, and saturation conditions.
%We are currently evaluating more complex scenarios that we outline in Section \ref{sec:future_work}.} \AzCom{Remove?! Can't really fit it in!}

In this section, the performance of the different CSMA protocols is evaluated through simulations. The presented results show the saturation throughput and fairness (Jain's Fairness Index (JFI) \cite{jain1984quantitative}) when the number of contenders increases. Additionally, the evolution of the probability that a slot contains a collision is plotted to provide further insights on the operation of each protocol, and specially whether the collision-free operation is reached.

%Additionally, the evolution of the probabilities that one slot is empty, contains a successful transmission or a collision, are plotted to provide further insights on the operation of each protocol.

To evaluate the four protocols, a network of $N$ nodes is considered, where each node is within the coverage area of the others. The channel does not introduce errors and the nodes are set to be in saturation (always have packets to transmit). The number of contenders, $N$, ranges from $2$ to $50$, and one thousand instances of the simulation are performed for each $N$. All plotted results show the 95\% confidence intervals.

A simulator of such scenario has been built, from scratch, using the C++ language and based on the COST (Component Oriented Simulation Toolkit) libraries \cite{yucesan2002cost}. The specific parameters of the IEEE 802.11n amendment \cite{IEEE80211n} are considered, and listed in Table \ref{Tbl:parameters}. In case packets are aggregated, default A-MPDU aggregation is considered. Further MAC-related parameters, as well as the code for the four CSMA protocols, are available online \cite{SanabriaSimulatorECA2012}.

\begin{table}[h!!!]
\centering
\begin{tabular}{c|c}
Parameter & Value \\
\hline
$CW_{\min}$ & $16$ slots \\
$S$ & 5 stages\\
% SIFS, DIFS & $9~\mu$s, $34~\mu$s \\
% $\sigma$ & $16~\mu$s \\
% $T_{\text{PHY}}$ & $32~\mu$s \\
% MAC header length (MH) & $288$ bits \\
% Service Field length (SF) & $16$ bits \\
% MPDU Delimiter length (MD) & $32$ bits \\
% Block ACK length ($L_{\text{BACK}}$) & $256$ bits \\
Data Packet Length & 12000 bits \\
Data Rate & $65$ Mbps\\
% Tail Bits (TB) & $6$ bits \\
% Data Bits Per Symbol $L_{\text{DBPS}}$ & $260$ bits/OFDM symbol \\
% OFDM Symbol Duration $T_{\text{s}}$ & $4~\mu$s \\
\hline
\end{tabular}

\caption{Parameters considered for the performance evaluation}\label{Tbl:parameters}

\end{table}

%A successful slot has a duration (in $\mu$s) of
%\begin{small}
%\begin{align}
% &T_{\text{slot,s}}(m)=\nonumber\\
% &T_{\text{PHY}} + \left \lceil \frac{\text{SF} + m(\text{MH} + L_{\text{DATA}}) + (m-1)\text{MD} + \text{TB} }{L_{\text{DBPS}}} \right \rceil T_{\text{s}} +\\
% &\text{SIFS}+ T_{\text{PHY}} + \left \lceil \frac{\text{SF} + L_{\text{BACK}} + \text{TB} }{L_{\text{DBPS}}} \right \rceil T_{\text{s}} +\text{DIFS}+\sigma \nonumber
%\end{align}
%\end{small}

\begin{figure}[ht!!!!!!!!]
\centering
\epsfig{file=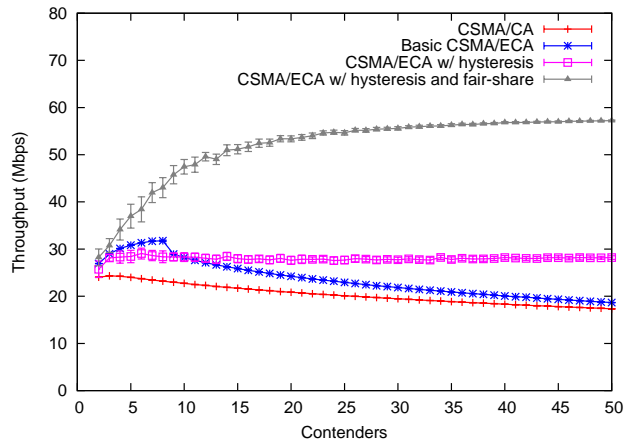,scale=0.45,angle=270}
\caption{Throughput}\label{Fig:throughput}
\end{figure}

In Figure \ref{Fig:throughput}, the throughput of all four schemes is plotted. As can be observed, the throughput for CSMA/CA decreases with the number of nodes, as the binary exponential backoff reduces the number of transmission attempts to keep the collisions low. The basic CSMA/ECA is able to achieve the collision-free operation if the number of nodes is lower than the deterministic cycle length, i.e., $N\leq \text{CW}_{\min}/2$ \AzCom{This time the $-1$ is not required, but it should be cycle length instead of backoff value, right?}. This is why in this figure, a phase transition is observed at $N=8$. When $N$ is larger than $\text{CW}_{\min}/2=8$, although the collision-free operation is not possible, the throughput obtained remains higher than in CSMA/CA. This is because, regardless of whether or not the collision-free operation is reached, in CSMA/ECA, a node always has a lower collision probability immediately after a successful transmission, without affecting the collision probability of other nodes.

\begin{figure}[ht!!!!!!!!]
\centering
\epsfig{file=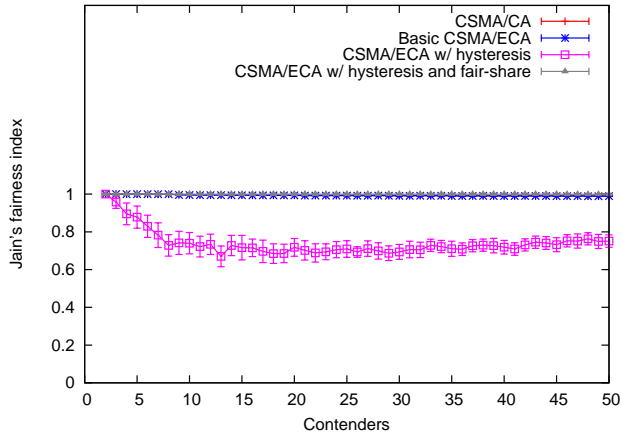,scale=0.45,angle=270}
\caption{Fairness}\label{Fig:fairness}
\end{figure}

\begin{figure}[ht!!!!!!!!]
\centering
\epsfig{file=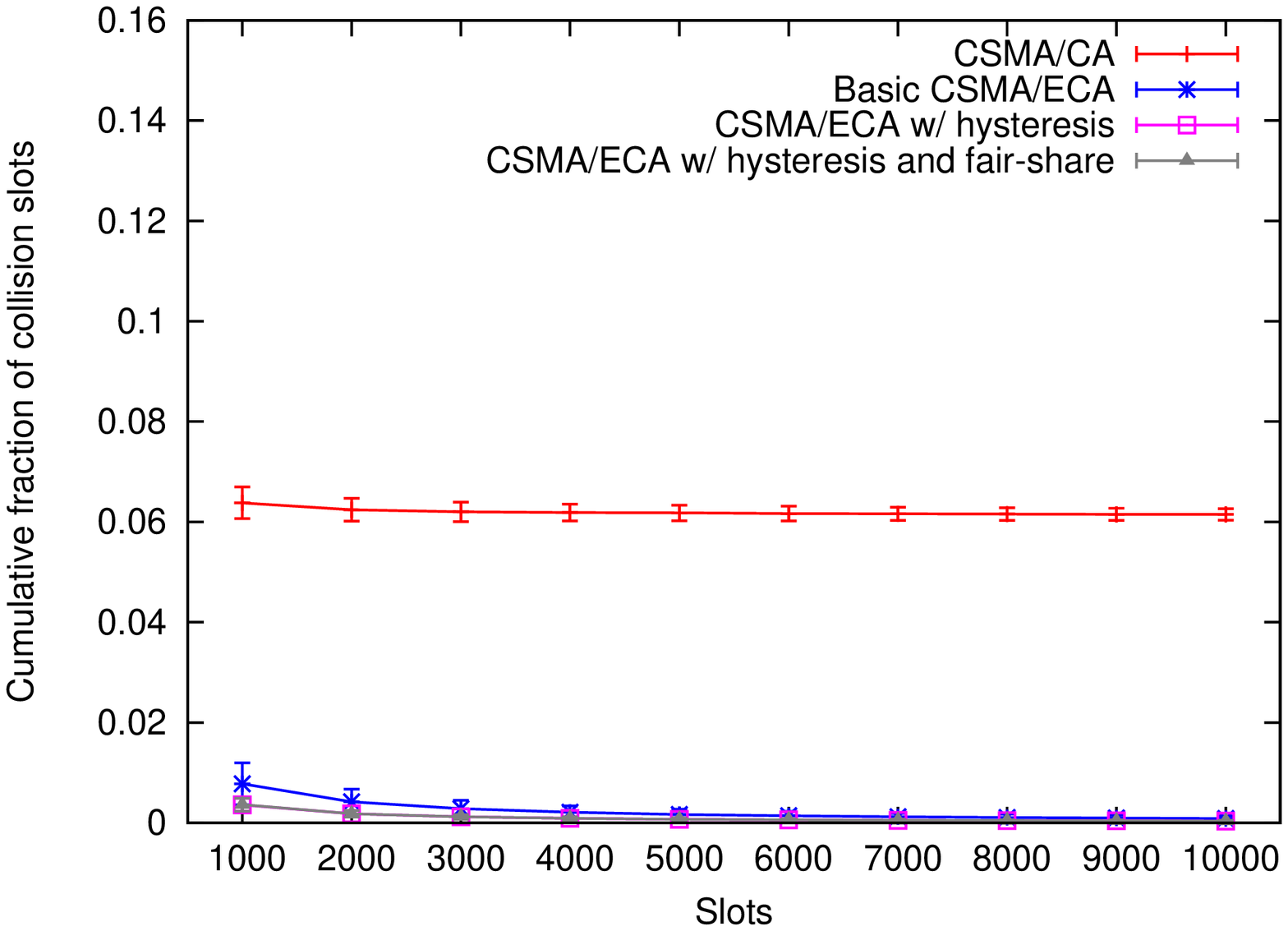,scale=0.45,angle=270}
\caption{Cumulative fraction of slots spent in collision for $N=6$ nodes}\label{Fig:Pc_6}
\end{figure}

\begin{figure}[ht!!!!!!!!]
\centering
\epsfig{file=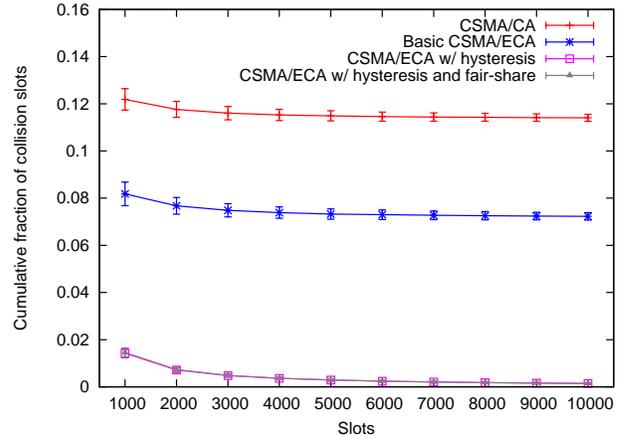,scale=0.45,angle=270}
\caption{Cumulative fraction of slots spent in collision for $N=12$ nodes}\label{Fig:Pc_12}
\end{figure}

Clearly, reaching a collision-free operation is highly desirable, but not always possible in the basic CSMA/ECA. CSMA/ECA with Hysteresis exactly addresses this issue by simply not resetting the backoff stage after a successful transmission. In other words, the deterministic backoff value ($2^s \text{CW}_{\min}/2-1$) is chosen based on the backoff stage, $s$, in which the transmission was successful. This way, a larger number of contenders can reach a collision-free operation (Figure \ref{Fig:throughput}), but at the expense of a lower throughput for $N\leq 8$ compared to the basic CSMA/ECA. This is because if collision-free operation can be reached with a lower deterministic backoff value, there will be fewer empty slots, hence higher channel access efficiency. Furthermore, CSMA/ECA with hysteresis has a lower long-term fairness (Figure \ref{Fig:fairness}) compared to the basic CSMA/ECA or even the legacy CSMA/CA protocol, as once the network has reached the collision-free state, nodes that use a large backoff stage will have fewer transmission opportunities than others.

To address the aforementioned fairness issue when hysteresis is used, CSMA/ECA with hysteresis and fair-share was introduced, \Az{whereby stations that access the channel less frequently use packet aggregation to send more (precisely, $2^s$) packets per attempt.} As observed in Figure \ref{Fig:throughput}, using aggregation we are able not only to provide fair access, but also to significantly increase the throughput. {The throughput increases with the number of nodes because for larger $N$, the collision-free operation is reached at higher backoff stages, which means more packets will be aggregated and transmitted at every attempt, thus improving the efficiency of the network.}

In Figures \ref{Fig:Pc_6} and \ref{Fig:Pc_12}, the evolution of the cumulative fraction of slots spent in collision is shown for $N = 6$ and $12$ nodes. As expected, for $N=6$, with all CSMA/ECA variants, collision-free operation is achieved (since $N<8$) and the probability that a slot contains a collision tends rapidly to zero. The collision-free operation is reached faster when hysteresis is used. When $N=12$, only the protocols using hysteresis reach collision-free operation.

\section{Conclusions and Future Work}
\label{sec:future_work}

In this paper we have described a suitable replacement of the prevalent CSMA/CA protocol used in the IEEE 802.11 standard. Compared to the CSMA/CA protocol, the proposed protocol, called CSMA/ECA with Hysteresis and Fair-Share, offers performance boost in terms of throughput, while preserving long-term fairness. In fact, under certain conditions, collision-free operation can be reached in a distributed and adaptive manner, even as the number of nodes in the network increases. Furthermore, this protocol is designed with backward compatibility in mind, so that it can operate with legacy nodes without degrading their performance. All these properties make the proposed protocol a good candidate to replace CSMA/CA in the upcoming revisions of the standard.

\Az{We have evaluated the protocols under discussion in a very simple scenario using ideal channel, ideal clocks, and in saturation conditions. Furthermore, all the participating stations use the same protocol and have a single traffic  class. Our proposed protocol is based on CSMA/ECA, which is shown \cite{he2009srb,barcelo2010fcc} to work well in non-ideal conditions, and, although not presented in this manuscript, simulation results confirm that CSMA/ECA with hysteresis and fair-share inherits the flexibility and robustness of its predecessors.

The proposed protocol has yet to be implemented in real prototypes. However, current IEEE 802.11 commercial network interfaces do not allow for setting the backoff to a deterministic value after successful transmissions, despite recent efforts on creating firmware that enables higher degree of manipulation  \cite{tinnirello2012wmp}. An alternative is using radio-frequency identification (RFID) devices, where there is a programmable hardware that allows for implementing arbitrary protocols. RFID uses also a contention protocol which is prone to collisions, and we believe it can be a first arena in which we will be able to develop working prototypes to validate some of the ideas proposed in the present paper.}

\section*{Acknowledgment}

The authors would like to thank the anonymous reviewers for their corrections, helpful comments and suggestions for future work.
This work has been partially supported by the Spanish Government under projects TEC2012-32354 and TEC2012-34642 (Plan Nacional I+D), CSD2008-00010 (Consolider-Ingenio) and by the Catalan Government (SGR2009\#00617 and SGR2009\#70).

% trigger a \newpage just before the given reference
% number - used to balance the columns on the last page
% adjust value as needed - may need to be readjusted if
% the document is modified later
%\IEEEtriggeratref{8}
% The "triggered" command can be changed if desired:
%\IEEEtriggercmd{\enlargethispage{-5in}}

% references section

% can use a bibliography generated by BibTeX as a .bbl file
% BibTeX documentation can be easily obtained at:
% http://www.ctan.org/tex-archive/biblio/bibtex/contrib/doc/
% The IEEEtran BibTeX style support page is at:
% http://www.michaelshell.org/tex/ieeetran/bibtex/
\bibliographystyle{IEEEtran}
% argument is your BibTeX string definitions and bibliography database(s)
\bibliography{IEEEabrv,my_bib}
%
% <OR> manually copy in the resultant .bbl file
% set second argument of \begin to the number of references
% (used to reserve space for the reference number labels box)
%\begin{thebibliography}{1}
%
%\bibitem{IEEEhowto:kopka}
%H.~Kopka and P.~W. Daly, \emph{A Guide to \LaTeX}, 3rd~ed.\hskip 1em plus
%  0.5em minus 0.4em\relax Harlow, England: Addison-Wesley, 1999.
%
%\end{thebibliography}

% that's all folks
\end{document}